# Depression Detection Based on Electroencephalography Using a Hybrid Deep Neural Network CNN-GRU and MRMR Feature Selection

Mohammad Reza Yousefi[1,2]*, Hajar Ismail Al-Tamimi [2], Amin Dehghani [3]

[1] Department of Electrical Engineering, Na.C., Islamic Azad University, Najafabad, Iran
[2] Digital Processing and Machine Vision Research Center, Na.C., Islamic Azad University, Najafabad, Iran
[3] Department of Psychological and Brain Sciences, Dartmouth College, Hanover, NH, USA
* mr-yousefi@iau.ac.ir

This study investigates the detection and classification of depressive and non-depressive states using deep learning approaches. Depression is a prevalent mental health disorder that substantially affects quality of life, and early diagnosis can greatly enhance treatment effectiveness and patient care. However, conventional diagnostic methods rely heavily on self-reported assessments, which are often subjective and may lack reliability. Consequently, there is a strong need for objective and accurate techniques to identify depressive states.

In this work, a deep learning–based framework is proposed for the early detection of depression using EEG signals. EEG data, which capture underlying brain activity and are not influenced by external behavioral factors, can reveal subtle neural changes associated with depression. The proposed approach combines convolutional neural networks (CNNs) and gated recurrent units (GRUs) to jointly extract spatial and temporal features from EEG recordings. The minimum redundancy maximum relevance (MRMR) algorithm is then applied to select the most informative features, followed by classification using a fully connected neural network.

The results demonstrate that the proposed model achieves high performance in accurately identifying depressive states, with an overall accuracy of 98.74%. By effectively integrating temporal and spatial information and employing optimized feature selection, this method shows strong potential as a reliable tool for clinical applications. Overall, the proposed framework not only enables accurate early detection of depression but also has the potential to support improved treatment strategies and patient outcomes.

# 1. Introduction

Major depressive disorder (MDD) is a serious mental health condition associated with substantial emotional, cognitive, and behavioral impairments [1–3]. Individuals with depression often experience persistent sadness, loss of interest or pleasure, and deficits in memory and attention. Long term emotional distress and cognitive dysfunction are common, and in severe cases, additional symptoms such as paranoia or hallucinations may occur [4]. Early diagnosis, when the disorder is still highly treatable, is therefore essential to improving outcomes and reducing long term complications [4,5].

Considerable research has been devoted to understanding the neural mechanisms underlying depression and sustained negative mood states. Traditionally, depression diagnosis has relied on clinical interviews and standardized rating scales administered by trained psychiatrists. Although effective, these methods are time consuming, subjective, and highly dependent on clinical expertise. In recent years, electroencephalography (EEG) has emerged as a promising objective tool for studying brain activity associated with depression [6–12], as well as other neurological and psychiatric conditions such as Alzheimer's disease (AD)[13–15], Parkinson's disease [16,17], epilepsy [18,19], and attention-deficit/hyperactivity disorder (ADHD) [20,21].

Despite growing awareness and improved diagnostic tools, many individuals with depression continue to experience delayed or inaccurate diagnoses, leading to prolonged suffering and increased risk of comorbidities. This challenge has motivated efforts to develop more objective, efficient, and reliable diagnostic approaches. Advances in sensor technology, mobile health platforms, and physiological signal analysis have created new opportunities for improving mental health assessment and early intervention.

EEG signals capture spontaneous electrical activity generated by cortical neurons via noninvasive scalp electrodes. Since the first recordings of human EEG, researchers have explored its potential to reveal links between neural function and mental disorders [22–24]. When combined with clinical and behavioral data, EEG can provide valuable insight into cognitive and emotional processes [25–31], making it a powerful modality for depression detection.

In this study, we propose a novel deep learning based framework for the early and accurate detection of depression using EEG signals. The proposed method integrates convolutional neural networks (CNN) and gated recurrent units (GRU) to capture both spatial and temporal characteristics of EEG data, leading to improved classification performance.

## 2. Literature Review

This section reviews related work on depression detection, which can be broadly categorized into three groups: approaches based on statistical feature extraction, conventional neural network models, and deep neural network architectures.

### 2.1 Methods Based on Statistical Features

Several studies have employed machine learning techniques using handcrafted EEG features for MDD detection. One framework extracted a diverse set of features, including statistical, spectral, wavelet based, functional connectivity, and nonlinear measures, from EEG signals. Sequential Backward Feature Selection (SBFS) was used to identify optimal feature subsets, and multiple classifiers were evaluated using a public EEG dataset comprising 34 MDD patients and 30 healthy controls [32].

In another study, alpha alpha one and alpha two, beta, delta, and theta band powers, along with theta asymmetry, were used as discriminative EEG features. Multi Cluster Feature Selection (MCFS) was applied to alpha sub bands combined with theta asymmetry. Statistical analysis revealed significant hemispheric differences in theta power among healthy individuals, whereas no such differences were observed in depressed participants [33].

Beyond EEG, other modalities have also been explored. A cross sectional online survey analyzed clinical and demographic factors such as stress, anxiety, and depression among healthcare workers and non clinical staff. Logistic regression revealed associations between occupational role and the likelihood of moderate to severe psychological symptoms after controlling for confounding variables [34]].

Text based approaches have also been investigated, with machine learning models trained to detect depression related patterns in social media posts, even when explicit references to depression were absent. Models trained on Twitter data were evaluated on external datasets from Facebook, Reddit, and electronic diaries, demonstrating varying degrees of generalizability [35].

## 2.2 Methods Based on Conventional Neural Networks

Conventional neural network models have been explored for depression detection, including hybrid approaches that fuse low-level acoustic descriptors with CNN-based spectral features. Applied to emotionally evocative reading data from 157 Chinese participants, these models achieved classification accuracies exceeding 82.7% [36]

A multimodal framework using three independent multilayer perceptron (MLP) networks in a feature fusion strategy was also introduced. This approach achieved competitive performance while maintaining relatively low computational complexity [37].

Actigraphy based methods have been explored as well. In one study, machine learning models were used to classify depression severity from wearable activity data, with XGBoost achieving the best performance. The results indicated that two days of actigraphy data were sufficient for effective feature extraction [38].

Behavioral and demographic data from 137 college students were used to train a support vector machine (SVM) for detecting persistent depressive disorder (PDD), achieving an accuracy of 89.4 percent. The findings highlighted demographic groups with elevated risk, supporting targeted early intervention strategies [39].

## 2.3 Methods Based on Deep Neural Networks

Deep learning approaches have demonstrated strong performance in depression detection across multiple data modalities. A generative long short term memory (LSTM) model was proposed to detect depression from textual data by capturing both semantic content and writing style patterns [40].

Another LSTM based recurrent neural network was applied to identify self reported depression symptoms in a large dataset of anonymous youth counseling messages. By incorporating medically informed structured features alongside textual data, the model achieved improved detection accuracy compared to word frequency based approaches [41].

For EEG based detection, a CNN based model known as DeprNet was introduced to classify EEG recordings into depressed and non depressed groups using PHQ-9 scores as ground truth labels. Both record wise and subject wise validation strategies demonstrated robust classification performance [42].

Overall, deep learning based methods have consistently outperformed traditional approaches in EEG based depression detection due to their ability to automatically learn complex and high level representations from raw data. Motivated by these findings, the present study employs a hybrid CNN-GRU architecture to capture both spatial and temporal dynamics in EEG signals.

## 3. Proposed Method

The proposed framework combines convolutional neural networks (CNN) and gated recurrent units (GRU) to model the spatial and temporal characteristics of electroencephalographic (EEG) signals for depression detection. The CNN component extracts localized spatial features by learning inter channel dependencies and electrode level correlations, while the GRU component captures sequential dependencies and long range temporal patterns within the EEG time series. By integrating these complementary architectures, the model effectively leverages both spatial feature extraction and temporal sequence modeling.

To further enhance performance and reduce computational complexity, the Minimum Redundancy Maximum Relevance (MRMR) algorithm is applied for feature selection. MRMR identifies features that are highly relevant to the classification task while minimizing redundancy among selected features. This process reduces dimensionality, suppresses noise, and improves generalization without compromising accuracy.

Overall, the proposed hybrid framework provides a robust and efficient approach for early depression detection, with strong potential for application in both clinical research and real world healthcare settings. A schematic overview of the proposed methodology is shown in Figure 1.

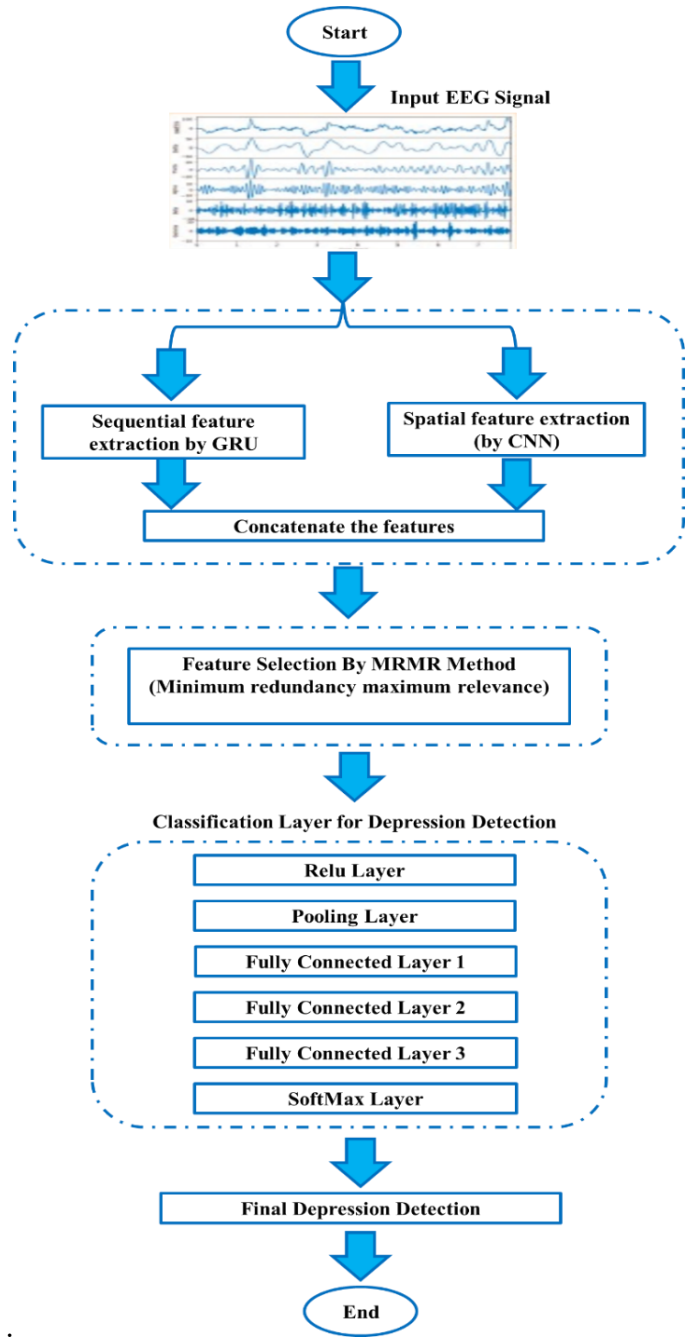

Figure 1: Diagram of the Proposed Method

## 3.1 Data Preprocessing

Raw EEG signals were denoised to remove electrical interference, motion artifacts, and other noise sources. Missing values were corrected and signals were normalized to ensure consistent scaling.

A bandpass filter (4.5–45 Hz) was applied to suppress irrelevant frequencies while preserving information relevant to depression-related brain activity.

## 3.2 Feature Extraction

Feature extraction integrated both spatial and temporal analyses of EEG signals. Spatial representations were learned using a convolutional neural network (CNN), which captured spatial correlations across EEG channels by processing channel-by-time matrices through successive convolutional, activation, pooling, and fully connected layers. In parallel, temporal dependencies were modeled using a gated recurrent unit (GRU) network, which leveraged update and reset gates to retain relevant sequential information over time. The outputs of the CNN and GRU branches were subsequently fused into a unified feature representation, enabling the model to exploit complementary spatial and temporal information for downstream analysis.

## 3.3 Feature Selection

The combined feature set was further refined using the Minimum Redundancy Maximum Relevance (mRMR) algorithm. This approach selects features that are highly informative with respect to the target class while minimizing redundancy among features, thereby reducing dimensionality, improving model generalization, and decreasing computational complexity.

## 3.4 Classification

The selected features were subsequently fed into a fully connected neural network for classification into depressive and healthy states. The network comprised multiple hidden layers with nonlinear activation functions and a Softmax output layer for binary classification. Model parameters were optimized via backpropagation using a cross-entropy loss function.

## 4 Evaluation of Results

System performance was evaluated following preprocessing, feature extraction, mRMR-based feature selection, and classification. Standard metrics, including accuracy, precision, recall, and F1 score, were computed to quantify the model's ability to discriminate between depressive and healthy states. The results indicate that the proposed hybrid CNN–GRU–mRMR–dense architecture provides effective EEG-based depression detection and demonstrates potential utility as an auxiliary tool in clinical settings.

## 4.1. Dataset

This study utilized the open-access MODMA dataset [20], which comprises EEG recordings from clinically diagnosed patients with major depressive disorder (MDD) and demographically matched healthy controls, with diagnoses confirmed by professional psychiatrists. For the present analysis, only EEG data were considered. Signals were collected under both resting-state and stimulation conditions using a wearable three-electrode EEG device. The dataset includes recordings from 24 individuals with MDD and 29 healthy controls, aged 16–52 years, along with associated demographic information and psychological assessment measures.

## 4.2. Evaluation Metrics

Model performance was evaluated using accuracy, precision, recall, and F1 score. Accuracy quantifies the overall proportion of correct predictions. Precision reflects the proportion of true positive cases among all positive predictions, whereas recall measures the proportion of true positive cases correctly identified. The F1 score, defined as the harmonic mean of precision and recall, provides a balanced assessment of performance, particularly in the presence of class imbalance.

## 4-3 Simulation Settings

A total of 53 EEG recordings were analyzed, including 29 healthy controls and 24 individuals with depression. Each recording was segmented into 10 equal-length epochs, yielding 530 EEG segments in total (290 healthy and 240 depressed). The dataset was randomly divided into training (70%, 371 segments) and testing (30%, 159 segments) sets to train the model and evaluate its performance on unseen data.

Feature extraction yielded 20 spatial features from the CNN and 100 temporal features from the GRU. These features were concatenated and subsequently refined using the mRMR algorithm, resulting in a reduced set of 30 informative features for classification.

The CNN convergence behavior is illustrated in Figure 2. Classification accuracy increased rapidly during training and stabilized after approximately 2000 iterations, while the corresponding loss curve decreased steadily and converged to a stable value. This convergence pattern suggests effective learning and indicates that the CNN was able to extract discriminative spatial features from EEG signals without evident overfitting.

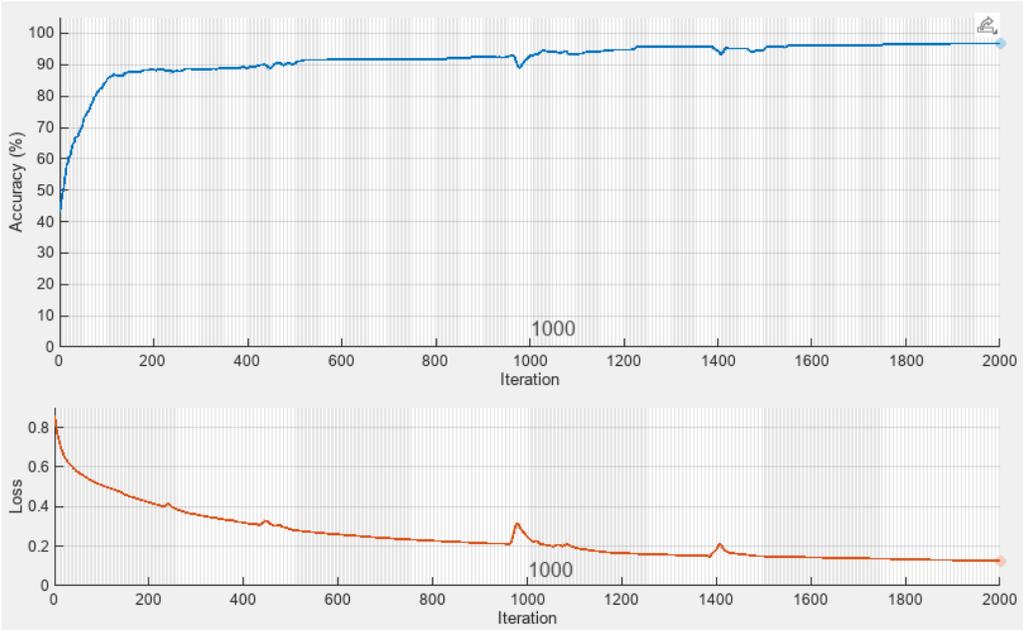

Figure 2: Convergence Curve of the Proposed CNN Network

The GRU convergence behavior is shown in Figure 3. Training accuracy increased steadily and stabilized at a high level midway through training, indicating effective learning of temporal dependencies in the EEG sequences. The corresponding loss curve decreased from approximately 0.4 to 0.1 before converging, reflecting progressive error reduction and stable optimization. Together, these trends suggest that the GRU effectively captured discriminative temporal patterns in the EEG data relevant to distinguishing depressive and healthy states.

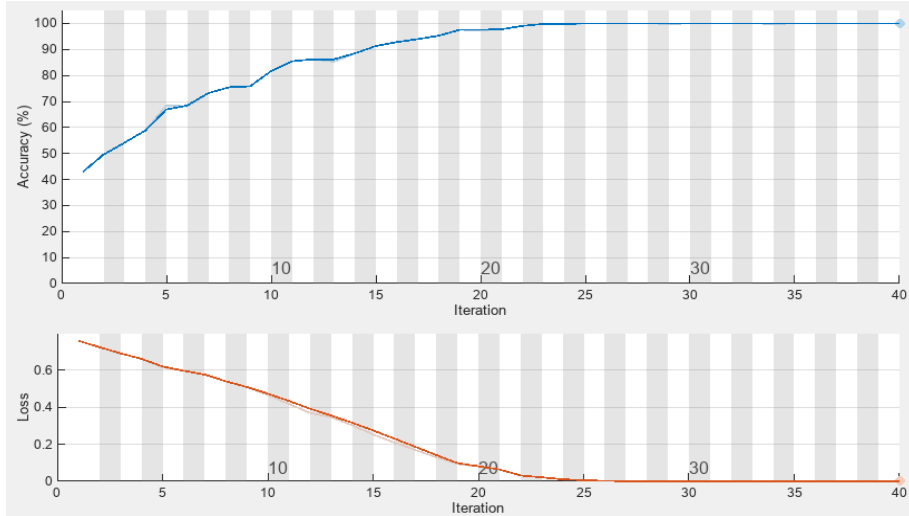

Figure 3: Convergence Curve of GRU

The convergence behavior of the final fully connected classifier is illustrated in Figure 4. Classification accuracy increased rapidly during training and plateaued at a high level within approximately 80–90 iterations, indicating stable optimization. The corresponding loss curve decreased from approximately 0.5 to below 0.05 before converging, reflecting progressive error reduction during training. Together, these results suggest that the classifier efficiently learned discriminative representations from the selected EEG features and achieved stable performance in depression classification.

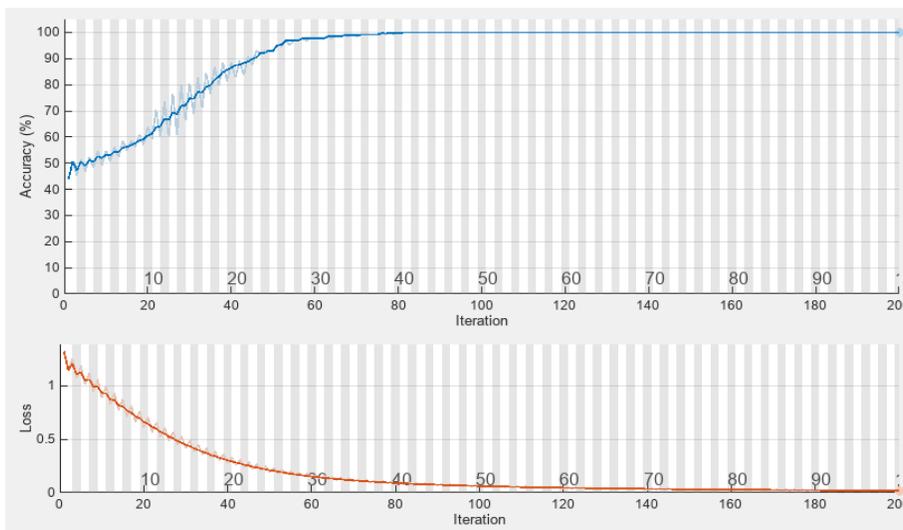

Figure 4: Convergence Curve of the Fully Connected Deep Classifier

## 4-4- Evaluation of Results

The confusion matrix (Figure 5) indicates that the model correctly classified 94 of 96 samples from the depressed group (97.2%) and all 63 samples from the healthy control group (100%). These results suggest strong discriminative performance across both classes and indicate robust generalization to previously unseen data. Together, the findings support the potential utility of the proposed model as an auxiliary tool for EEG-based depression assessment.

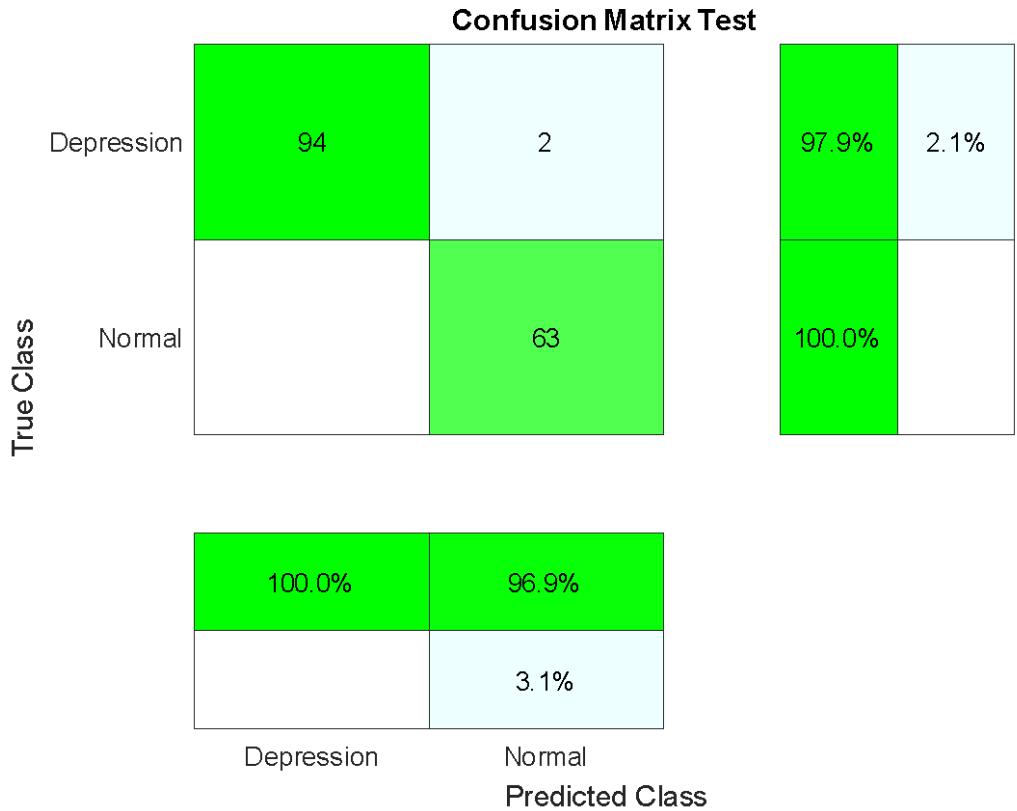

Figure 5: Confusion Matrix on Test Data

The regression analysis (Figure 6) demonstrates a strong correspondence between predicted and observed values, with the best-fit line exhibiting a slope of approximately 0.98. This indicates a high level of agreement between model predictions and ground truth. The mean prediction error was 0.042, reflecting low overall estimation error. Together, these results suggest accurate model fitting and stable generalization performance on unseen data, supporting the potential utility of the proposed approach for depression-related prediction tasks.

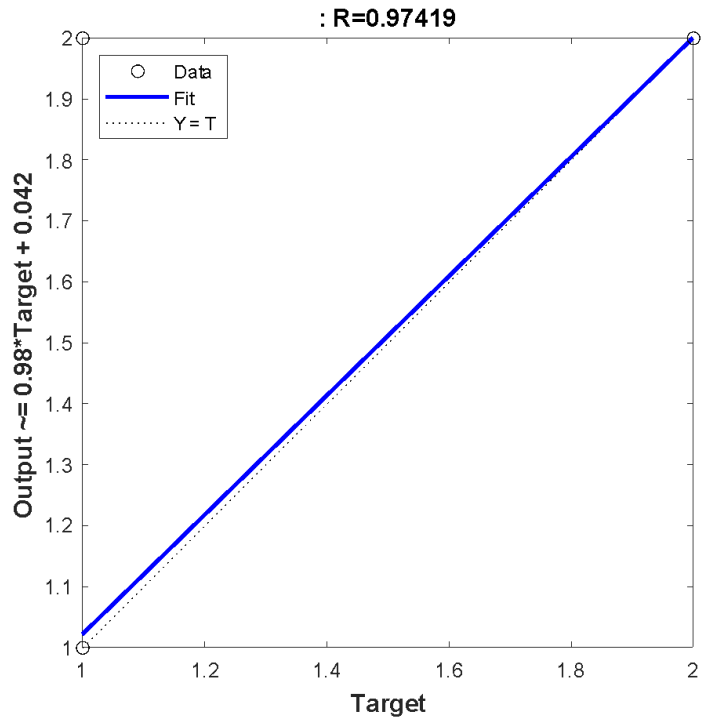

Figure 6: Regression Plot on Test Data

The receiver operating characteristic (ROC) curves shown in Figure 7 demonstrate strong discriminative performance between depressed and healthy classes. The curves approach the upper-left region of the ROC space, indicating high true positive rates with low false positive rates across classification thresholds. The area under the curve (AUC) was close to 1, reflecting robust predictive performance. These findings support the effectiveness of the proposed model for EEG-based depression classification and its potential utility as an auxiliary clinical assessment tool.

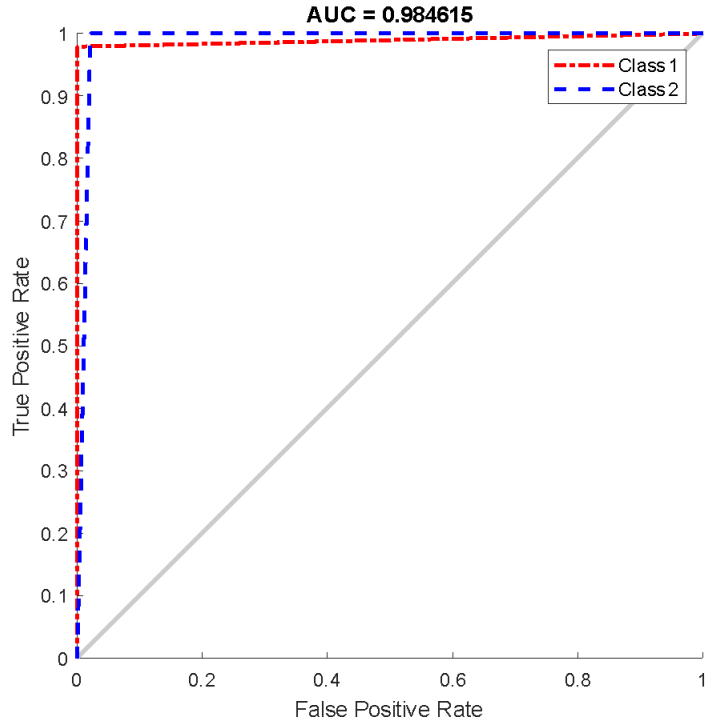

Figure 7: ROC Curve on Test Data

The quantitative performance metrics summarized in Figure 8 further demonstrate the strong classification performance of the proposed model. Specifically, the model achieved an accuracy of 96.74%, a precision of 97.91%, a recall of 100%, and an F1 score of 98.94%. These results indicate robust discrimination between depressed and healthy states and suggest stable and reliable performance, supporting the potential utility of the model as an auxiliary tool for EEG-based depression assessment.

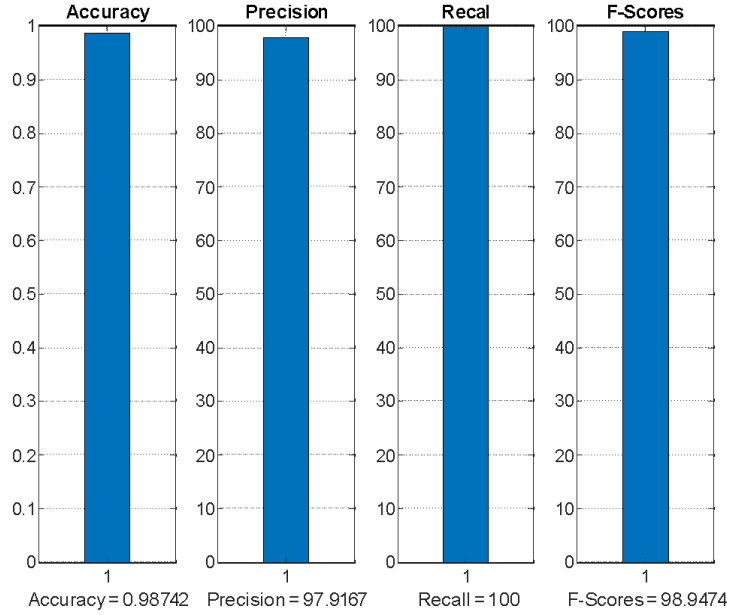

Figure 8: Numerical Metrics on Test Data

**4-5- Results Comparison**

Table 1 compares the classification accuracy of the proposed CNN–GRU–mRMR–Dense model with several deep learning architectures, including CNN-, GRU-, LSTM-, and ResNet-50–based configurations. The accuracy of the proposed method was averaged across 30 independent simulation runs, yielding a mean accuracy of 98.42%. This performance exceeds that of the compared models, indicating that the hybrid architecture effectively integrates spatial, temporal, and feature-selection components for EEG-based depression detection.

Table 1: Comparison of the Proposed Method with Other Methods

| Method | Accuracy (%) |
|---|---|
| CNN + GRU [43] | 89.63 |
| CNN [44] | 91.01 |
| ResNet-50 + LSTM [45] | 90.02 |
| CNN Network [46] | 97.00 |
| Proposed Method (CNN-GRU-MRMR-Dense) | 98.42 |

The strong classification performance of the proposed method can be attributed to the complementary integration of CNN and GRU components, which enables simultaneous extraction of spatial and temporal features from EEG signals. The inclusion of mRMR-based feature selection further enhances performance by retaining the most informative and nonredundant features. Together, these elements contribute to accurate and reliable EEG-based depression classification.

## 5. Discussion

This study introduced a hybrid EEG-based framework for automated depression detection that integrates convolutional neural networks for spatial feature extraction, gated recurrent units for temporal modeling, and minimum redundancy maximum relevance feature selection to optimize classification performance. By jointly modeling spatial and temporal characteristics of EEG signals, the proposed CNN–GRU–mRMR architecture achieved a mean classification accuracy of 98.42%, outperforming several established deep learning approaches. These findings underscore the importance of capturing complementary neural information across multiple representational domains when modeling depression-related brain activity.

The strong performance of the proposed model highlights the value of combining spatial and temporal representations of EEG data. CNN-based feature extraction effectively captured inter-channel relationships and electrode-level spatial patterns, while the GRU component modeled long-range temporal dependencies inherent to EEG time series. The integration of these complementary architectures enabled more comprehensive characterization of neural dynamics associated with depressive states than single-model approaches. In addition, the use of mRMR feature selection reduced redundancy and emphasized the most informative features, improving both computational efficiency and generalization performance.

From a clinical perspective, the high recall achieved by the proposed method is particularly notable, as it indicates effective identification of depressive cases, a critical requirement for screening and early detection tools. High precision further suggests robustness against false-positive classifications, reducing the likelihood of mislabeling healthy individuals. Together with the strong ROC performance and stable convergence behavior, these results support the feasibility of EEG-based deep learning models as objective adjuncts to traditional diagnostic assessments.

Beyond diagnostic accuracy, objective EEG-derived biomarkers have important implications for treatment development and evaluation. Noninvasive neuromodulation techniques, such as transcranial direct current stimulation (tDCS) or transcranial magnetic stimulation, have demonstrated the ability to modulate cortical excitability and influence large-scale brain networks implicated in affective and cognitive processing [47–51]. Prior work has shown that targeted tDCS can reliably alter neural and behavioral responses, highlighting the sensitivity of distributed brain networks to externally applied stimulation and the importance of individualized neural characterization [52–56]. In this context, EEG-based depression detection frameworks may serve not only as diagnostic tools but also as complementary biomarkers for monitoring treatment effects, stratifying patients, and informing personalized neuromodulation strategies.

Despite the promising performance of the proposed framework, several limitations should be acknowledged. The dataset used in this study is relatively small and limited to three EEG channels, which may constrain generalizability. Future studies should evaluate the proposed method on larger, more diverse cohorts and extend validation to high-density EEG recordings. Additionally, external validation and real-time implementation will be essential to assess robustness in clinical environments.

Overall, the proposed hybrid CNN–GRU–mRMR framework represents a step toward objective, scalable, and clinically relevant EEG-based depression assessment. With further validation and integration into treatment-monitoring paradigms, such approaches may contribute to more precise diagnosis, personalized intervention planning, and improved outcomes for individuals with depression.

## 6. Conclusion

This study presented a hybrid deep learning framework that integrates convolutional neural networks, gated recurrent units, and minimum redundancy maximum relevance feature selection for automated depression detection from EEG signals. By jointly modeling spatial and temporal characteristics of neural activity and emphasizing informative, nonredundant features, the proposed approach achieved strong classification performance, demonstrating its effectiveness for EEG-based depression assessment.

The proposed framework highlights the importance of combining complementary representations of EEG data to capture complex neural dynamics associated with depressive states. Spatial feature extraction enabled characterization of inter-channel relationships, while temporal modeling captured sequential dependencies inherent to EEG time series. The integration of feature selection further enhanced model efficiency and robustness, supporting reliable discrimination between depressive and healthy conditions.

Beyond diagnostic performance, EEG-based deep learning models such as the one introduced here have broader implications for mental health research and clinical practice. Objective neural markers derived from EEG may support scalable screening approaches and facilitate monitoring of disease-related neural changes over time. In addition, as noninvasive neuromodulation techniques increasingly target distributed brain networks, EEG-informed models may contribute to personalized intervention strategies by providing quantitative measures of neural state and responsiveness.

Overall, the proposed CNN–GRU–MRMR framework represents a promising step toward objective, noninvasive, and data-driven assessment of depression. By leveraging advances in deep learning and neurophysiological signal analysis, this approach has the potential to support improved diagnosis, treatment evaluation, and precision-oriented mental health care.

# References

[1] McIntyre RS, Cha DS. Cognitive Impairment in Major Depressive Disorder: Clinical Relevance, Biological Substrates, and Treatment Opportunities. Cambridge University Press; 2016.
[2] McIntyre RS, Cha DS, Soczynska JK. Cognition in Major Depressive Disorder. Oxford University Press, USA; 2013.
[3] Young AH, Harmer C. Cognition in Mood Disorders. Frontiers Media SA; 2020.
[4] Safayari A, Bolhasani H. Depression diagnosis by deep learning using EEG signals: A systematic review. Med Nov Technol Devices 2021;12:100102.
[5] Sarkar A, Singh A, Chakraborty R. A deep learning-based comparative study to track mental depression from EEG data. Neurosci Inform 2022;2:100039.
[6] Ay B, Yildirim O, Talo M, Baloglu UB, Aydin G, Puthankattil SD, et al. Automated Depression Detection Using Deep Representation and Sequence Learning with EEG Signals. J Med Syst 2019;43:205.
[7] Dehghani A, Soltanian-Zadeh H, Hossein-Zadeh G-A. Neural modulation enhancement using connectivity-based EEG neurofeedback with simultaneous fMRI for emotion regulation. Neuroimage 2023;279:120320.
[8] Dehghani A, Soltanian-Zadeh H, Hossein-Zadeh G-A. Probing fMRI brain connectivity and activity changes during emotion regulation by EEG neurofeedback. Front Hum Neurosci 2022;16:988890.
[9] Ebrahimzadeh E, Dehghani A, Asgarinejad M, Soltanian-Zadeh H. Non-linear processing and reinforcement learning to predict rTMS treatment response in depression. Psychiatry Res Neuroimaging 2023;337:111764.
[10] Mosayebi R, Dehghani A, Hossein-Zadeh G-A. Dynamic functional connectivity estimation for neurofeedback emotion regulation paradigm with simultaneous EEG-fMRI analysis. Front Hum Neurosci 2022;16:933538.
[11] Yousefi MR, Dehghani A, Taghaavifar H. Enhancing the accuracy of electroencephalogram-based emotion recognition through Long Short-Term Memory recurrent deep neural networks. Frontiers in Human Neuroscience 2023.
[12] Ebrahimzadeh E, Sadjadi SM, Asgarinejad M, Dehghani A, Rajabion L, Soltanian-Zadeh H. Neuroenhancement by repetitive transcranial magnetic stimulation (rTMS) on DLPFC in healthy adults. Cogn Neurodyn 2025;19:34.
[13] Kulkarni N, Bairagi V. EEG-Based Diagnosis of Alzheimer Disease: A Review and Novel Approaches for Feature Extraction and Classification Techniques. Academic Press; 2018.
[14] Jeong J. EEG dynamics in patients with Alzheimer's disease. Clin Neurophysiol 2004;115:1490–505.
[15] Moretti DV, Babiloni C, Binetti G, Cassetta E, Dal Forno G, Ferreric F, et al. Individual analysis of EEG frequency and band power in mild Alzheimer's disease. Clin Neurophysiol 2004;115:299–308.
[16] Geraedts VJ, Boon LI, Marinus J, Gouw AA, van Hilten JJ, Stam CJ, et al. Clinical correlates of quantitative EEG in Parkinson disease: A systematic review. Neurology 2018;91:871–83.
[17] Bera S, Geem ZW, Cho Y-I, Singh PK. A Comparative Study of Machine Learning and Deep Learning Models for Automatic Parkinson's Disease Detection from Electroencephalogram Signals. Diagnostics (Basel) 2025;15. https://doi.org/10.3390/diagnostics15060773.
[18] Ebrahimzadeh E, Shams M, Seraji M, Sadjadi SM, Rajabion L, Soltanian-Zadeh H. Localizing epileptic foci using simultaneous EEG-fMRI recording: Template component cross-correlation. Front Neurol 2021;12:695997.
[19] Ebrahimzadeh E, Soltanian-Zadeh H, Araabi BN, Fesharaki SSH, Habibabadi JM. Component-related BOLD response to localize epileptic focus using simultaneous EEG-fMRI recordings at 3T. J Neurosci Methods 2019;322:34–49.
[20] Yousefi MR, Khanahmadi N, Dehghani A. Utilizing Phase Locking Value to Determine Neurofeedback Treatment Responsiveness in Attention Deficit Hyperactivity Disorder. J Integr Neurosci 2024;23:121.

[21] Arns M, Conners CK, Kraemer HC. A decade of EEG Theta/Beta Ratio Research in ADHD: a meta-analysis. J Atten Disord 2013;17:374–83.
[22] Hawes MT, Szenczy AK, Klein DN, Hajcak G, Nelson BD. Increases in depression and anxiety symptoms in adolescents and young adults during the COVID-19 pandemic. Psychol Med 2022;52:3222–30.
[23] Rivera MJ, Teruel MA, Maté A, Trujillo J. Diagnosis and prognosis of mental disorders by means of EEG and deep learning: a systematic mapping study. Artif Intell Rev 2022;55:1209–51.
[24] Luján MÁ, Mateo Sotos J, Torres A, Santos JL, Quevedo O, Borja AL. Mental Disorder Diagnosis from EEG Signals Employing Automated Leaning Procedures Based on Radial Basis Functions. J Med Biol Eng 2022;42:853–9.
[25] Li Y, Shen Y, Fan X, Huang X, Yu H, Zhao G, et al. A novel EEG-based major depressive disorder detection framework with two-stage feature selection. BMC Med Inform Decis Mak 2022;22:209.
[26] Dehghani A, Soltanian-Zadeh H, Hossein-Zadeh G-A. Global data-driven analysis of brain connectivity during emotion regulation by electroencephalography neurofeedback. Brain Connect 2020;10:302–15.
[27] Dehghani A, Soltanian-Zadeh H, Hossein-Zadeh G-A. EEG coherence pattern through recalling positive autobiographical memories and neurofeedback. 2021 28th National and 6th International Iranian Conference on Biomedical Engineering (ICBME), IEEE; 2021. https://doi.org/10.1109/icbme54433.2021.9750357.
[28] Ray WJ, Cole HW. EEG activity during cognitive processing: influence of attentional factors. Int J Psychophysiol 1985;3:43–8.
[29] Ray WJ, Cole HW. EEG alpha activity reflects attentional demands, and beta activity reflects emotional and cognitive processes. Science 1985;228:750–2.
[30] Kim M-K, Kim M, Oh E, Kim S-P. A review on the computational methods for emotional state estimation from the human EEG. Comput Math Methods Med 2013;2013:573734.
[31] Gkintoni E, Halkiopoulos C. Mapping EEG Metrics to Human Affective and Cognitive Models: An Interdisciplinary Scoping Review from a Cognitive Neuroscience Perspective. Biomimetics (Basel) 2025;10. https://doi.org/10.3390/biomimetics10110730.
[32] Movahed RA, Jahromi GP, Shahyad S, Meftahi GH. A major depressive disorder classification framework based on EEG signals using statistical, spectral, wavelet, functional connectivity, and nonlinear analysis. J Neurosci Methods 2021;358:109209.
[33] Mahato S, Paul S. Classification of Depression Patients and Normal Subjects Based on Electroencephalogram (EEG) Signal Using Alpha Power and Theta Asymmetry. J Med Syst 2019;44:28.
[34] Nemesure MD, Heinz MV, Huang R, Jacobson NC. Predictive modeling of depression and anxiety using electronic health records and a novel machine learning approach with artificial intelligence. Sci Rep 2021;11:1980.
[35] Chiong R, Budhi GS, Dhakal S, Chiong F. A textual-based featuring approach for depression detection using machine learning classifiers and social media texts. Comput Biol Med 2021;135:104499.
[36] Wang C, Zhang X, Sun J, Zhang Z, Liang Z, Yu Z, et al. Detection of major depressive disorder in adolescents based on textual and acoustic features. J Affect Disord 2026;396:120795.
[37] Sun H, Wang H, Liu J, Chen Y-W, Lin L. CubeMLP: An MLP-based model for multimodal sentiment analysis and depression estimation. Proceedings of the 30th ACM International Conference on Multimedia, New York, NY, USA: ACM; 2022. https://doi.org/10.1145/3503161.3548025.
[38] Choi J-G, Ko I, Han S. Depression level classification using machine learning classifiers based on actigraphy data. IEEE Access 2021;9:116622–46.
[39] Upadhyay DK, Mohapatra S, Singh NK. An early assessment of Persistent Depression Disorder using machine learning algorithm. Multimed Tools Appl 2023;83:49149–71.
[40] Amanat A, Rizwan M, Javed AR, Abdelhaq M, Alsaqour R, Pandya S, et al. Deep learning for

depression detection from textual data. Electronics (Basel) 2022;11:676.
[41] Uddin MZ, Dysthe KK, Følstad A, Brandtzaeg PB. Deep learning for prediction of depressive symptoms in a large textual dataset. Neural Comput Appl 2022;34:721–44.
[42] Seal A, Bajpai R, Agnihotri J, Yazidi A, Herrera-Viedma E, Krejcar O. DeprNet: A deep convolution neural network framework for detecting depression using EEG. IEEE Trans Instrum Meas 2021;70:1–13.
[43] Liu W, Jia K, Wang Z, Ma Z. A Depression Prediction Algorithm Based on Spatiotemporal Feature of EEG Signal. Brain Sci 2022;12. https://doi.org/10.3390/brainsci12050630.
[44] Wang B, Kang Y, Huo D, Feng G, Zhang J, Li J. EEG diagnosis of depression based on multi-channel data fusion and clipping augmentation and convolutional neural network. Front Physiol 2022;13:1029298.
[45] Uyulan C, de la Salle S, Erguzel TT, Lynn E, Blier P, Knott V, et al. Depression Diagnosis Modeling With Advanced Computational Methods: Frequency-Domain eMVAR and Deep Learning. Clin EEG Neurosci 2022;53:24–36.
[46] Ksibi A, Zakariah M, Menzli LJ, Saidani O, Almuqren L, Hanafieh RAM. Electroencephalography-Based Depression Detection Using Multiple Machine Learning Techniques. Diagnostics (Basel) 2023;13. https://doi.org/10.3390/diagnostics13101779.
[47] Boggio PS, Zaghi S, Fregni F. Modulation of emotions associated with images of human pain using anodal transcranial direct current stimulation (tDCS). Neuropsychologia 2009;47:212–7.
[48] Dehghani A, Bango C, Murphy EK, Halter RJ, Wager TD. Independent effects of transcranial direct current stimulation and social influence on pain. Pain 2025;166:87–98.
[49] Nitsche MA, Koschack J, Pohlers H, Hullemann S, Paulus W, Happe S. Effects of frontal transcranial direct current stimulation on emotional state and processing in healthy humans. Front Psychiatry 2012;3:58.
[50] Vonck S, Swinnen SP, Wenderoth N, Alaerts K. Effects of Transcranial Direct Current Stimulation on the Recognition of Bodily Emotions from Point-Light Displays. Front Hum Neurosci 2015;9:438.
[51] Dehghani A, Gantz DM, Murphy EK, Nitsche MA, Halter RJ, Wager TD. Transcranial direct current stimulation of primary motor cortex reduces thermal pain. Pain 2025. https://doi.org/10.1097/j.pain.0000000000003851.
[52] Nitsche MA, Paulus W. Excitability changes induced in the human motor cortex by weak transcranial direct current stimulation. J Physiol 2000;527 Pt 3:633–9.
[53] Jacobson L, Koslowsky M, Lavidor M. tDCS polarity effects in motor and cognitive domains: a meta-analytical review. Exp Brain Res 2012;216:1–10.
[54] Polanía R, Nitsche MA, Paulus W. Modulating functional connectivity patterns and topological functional organization of the human brain with transcranial direct current stimulation. Hum Brain Mapp 2011;32:1236–49.
[55] Huang Y, Liu AA, Lafon B, Friedman D, Dayan M, Wang X, et al. Correction: Measurements and models of electric fields in the human brain during transcranial electric stimulation. Elife 2018;7. https://doi.org/10.7554/eLife.35178.
[56] Krause B, Cohen Kadosh R. Not all brains are created equal: the relevance of individual differences in responsiveness to transcranial electrical stimulation. Front Syst Neurosci 2014;8:25.